\documentclass[twocolumn,preprintnumbers,amsmath,amssymb]{revtex4}
\usepackage{amssymb}
\usepackage{latexsym}
\begin{document}

\title{$4$-index theory of gravity and its relation with the violation of the energy-momentum conservation law}

\author{H. Moradpour$^1$\footnote{h.moradpour@riaam.ac.ir}, I. Licata$^{1,2,3}$\footnote{ignazio.licata3@gmail.com}, C. Corda$^{1,4}$\footnote{cordac.galilei@gmail.com
}, Ines G. Salako$^{5}$\footnote{inessalako@gmail.com}}
\address{$^1$ Research Institute
for Astronomy and Astrophysics of Maragha (RIAAM), P.O. Box
55134-441, Maragha, Iran\\
$^2$ ISEM, Inst. for Scientific Methodology, PA, Italy\\
$^3$ School of Advanced International Studies on Applied
Theoretical and Non Linear Methodologies in
Physics, Bari (Italy)\\
$^4$International Institute for Applicable Mathematics and Information Sciences, Adarshnagar, Hyderabad 500063 (India)\\
$^5$ Institut de Math\'ematiques et de Sciences Physiques (IMSP)
01 BP 613 Porto-Novo, B\'enin}

\begin{abstract}
Recently, a $4$-index generalization of the Einstein theory is
proposed by Moulin \cite{4index}. Using this method, we find the
most general $2$-index field equations derivable from the
Einstein-Hilbert action. The application of Newtonian limit, the
role of gravitational coupling constant and the effects of the
properties of ordinary energy-momentum tensor in obtaining a
$4$-index gravity theory have been studied. We also address the
results of building Weyl free $4$-index gravity theory. Our study
displays that both the Einstein and Rastall theories can be
obtained as the subclasses of a $4$-index gravity theory which
shows the power of $4$-index method in unifying various
gravitational theories. It is also obtained that the violation of
the energy-momentum conservation law may be allowed in $4$-index
gravity theory, and moreover, the contraction of $4$-index theory
generally admits a non-minimal coupling between geometry and
matter field in the Rastall way. This study also shows that,
unlike the Einstein case, the gravitational coupling constant of
$4$-index Rastall theory generally differs from that of the
ordinary $2$-index Rastall theory.
\end{abstract}
\maketitle


\section{Introduction\label{Intr}}
In Riemannian geometry, the geometrical information of a manifold
is encoded into the forth-order Riemann tensor, while the general
relativity (GR) includes the second rank divergence-less tensors
\cite{4not,pois}. Remember that the Einstein tensor is a
combination of the Ricci tensor and its scalar constructed by
contracting the Riemann tensor.

Based on the Einstein hypothesis, whenever the spacetime is curved
in the presence of an energy source, the amount of curvature is
determinable by the total energy-momentum tensor of source filling
the background. In fact, whenever the GR field equations are
solved, then metric and thus the Riemann tensor are also
determined, and we can find more information about spacetime by
studying the Riemann tensor and its evolution. This approach is in
fact a secondary way to study the Riemann tensor in which we do
not know anything about the probable constraints on this tensor
and its evolution. It means that GR does not directly say us
anything about the evolution of this curvature. Thus, a proper
theory should help us in obtaining the Riemann tensor and its
evolution in a direct way and without any intermediary.

Theoretically, another primary way to directly find some
information about the evolution of geometry may also be to
establish a relation between the changes of geometry and that of
the energy-momentum source filling it. In GR, at first glance,
such a relation is exist between the derivatives of Ricci scalar
(or Ricci tensor) and the trace of energy momentum tensor. But,
this is a secondary relation directly obtained from the field
equations which are the fundamental and basic equations in GR.
Therefore, another basic assumptions are needed to build a theory
which lets us directly study the evolution of geometry.

On the other hand, if one relaxes the energy-momentum conservation
law by considering a mutual interaction between the geometry and
matter field \cite{rastall,cmc,cmc1,cmc2,genras,rl3,rl4,rl5}, then
it can at least theoretically provide for us an independent basic
to understand this issue that how the geometry is curved and its
curvature evolves. Rastall theory is a leading work in this sense
\cite{rastall} which proposes a fundamental basic equation between
the evolution of geometry and the changes of energy source, in a
way it is ahead of its corresponding field equations. In fact,
field equations in Rastall hypothesis, a modified GR, are
secondary equations. Although this is an old theory, its
Lagrangian is still under debate \cite{rl3,rl4,rl5,rl1,rl2}. It is
worthwhile mentioning that recent observation indicate that the
gravitational waves are propagated with the light velocity
\cite{prln}, a result respected by both the Einstein and Rastall
gravity \cite{4not,mcl}, and severely restricts gravitational
theories as well as dark energy models \cite{prln}. Rastall
gravity is inserted in the more general framework of extended
gravity [19 - 24], which is today considered an intriguing
tapestry to challenge the big puzzles of the standard model of
cosmology, starting from the famous dark energy
\cite{Riess,Riess2} and dark matter \cite{Clowe,Navarro} problems.
We emphasize that all of the potential alternatives to general
relativity must be viable theories. This means that alternative
theories must be consistent with Einstein's equivalence principle
and, in turn, they must be metric theories \cite{will}. In fact,
Einstein's equivalence principle is today supported by an
unchallengeable empirical evidence \cite{will}, being considered
at the level of an important law of Nature. Another request is
that alternative theories must pass the solar system tests. As a
consequence, deviations of extended theories from standard general
relativity must be weak \cite{Corda,Corda2}. Remarkably, the
nascent gravitational wave astronomy \cite{LIGO} could be a useful
tool in order to discriminate among Einstein gravity, Rastall
gravity and other potential alternative theories
\cite{mcl,Corda,Corda2}. In fact, important differences between
general relativity and extended theories can be pointed out in the
linearized theory of gravity \cite{Corda,Corda2}.

Recently, a $4$-index generalization of general relativity has
been introduced to relate the Riemann tensor to the
energy-momentum sources filling the spacetime \cite{4index}.
Therefore, it is a theory may directly help us in modelling the
evolution of geometry and thus its curvature. In this approach,
the gravitational field contribution to the total energy-momentum
tensor is separated from other sources, and it is related to the
Weyl tensor defined as \cite{pois}

\begin{eqnarray}\label{weyl}
&&C_{ijkl}=R_{ijkl}-\frac{1}{n-1}(g_{ijkp}R^p_{\ l}+g_{ijpl}R^p_{\
k})\nonumber\\&&+\frac{1}{n(n-1)}g_{ijkl}R,
\end{eqnarray}

\noindent and is zero for conformal flat spacetimes.

Therefore, the view of Ref~\cite{4index} claims that the
gravitational field has not any effects in conformal flat
spacetimes which are indeed curved. If we accept the Einstein idea
that spacetime is curved by energy sources, then this property of
Weyl tensor will establishes an inconsistency with his idea. It is
because conformal flat spacetimes are in fact curved whereas their
Weyl tensor is zero. In fact, if physics is formulated in terms of
this tensor, then some information will be disappeared whenever we
face with conformal flat spacetimes such as the FRW geometry. It
means that, in this situation, we should probably establish
another set of equations to get the missed information meaning
that the theory is incomplete. Briefly, this tensor does not has a
unique behavior in front of the existence of curvature. Hence,
although Weyl tensor includes some information about the geometry,
due to its dual behavior against the existence of curvature, one
may argue that a true $4$-index generalization of Einstein theory
should not include Weyl tensor.

Based on the above argument, we are going to show that the
$4$-index approach, introduced in \cite{4index}, may provide a
Lagrangian description for the Rastall theory. We are also
interested in studying the role of gravitational coupling
constant, the application of Newtonian limit and the results of
building a Weyl tensor free gravitational field equations in this
approach.

The paper is organized as follows. In the next section, after
reviewing the lagrangian formalism in both $2$ and $4$-index
notations, and addressing the role of gravitational coupling
constant, we build the general $4$-index gravitational field
equations extractable by the Einstein-Hilbert Lagrangian. The
conditions required for obtaining Einstein and Rastall theory have
also been studied. Section~($\textmd{III}$) includes our surgery
on the obtainable $2$-index theories from this approach by
generalizing the gravitational action used in the second section.
The last section is devoted to a summary and concluding remarks.

\section{Action, its variation, $2$-and $4$-index theories}

Before focusing on our main aim, we review some features of the
Lagrangian formalism of general relativity in ($n+1$)-dimension.

\subsection{($n+1$)-dimensional general relativity}

The Einstein-Hilbert action is

\begin{equation}\label{action}
I=I_G+I_m,
\end{equation}

\noindent in which

\begin{equation}\label{lg}
I_G=-\frac{1}{2\kappa_n}\int R \sqrt{-g}\ d^{n+1}x,
\end{equation}

\noindent is the gravitational action, $I_m$ denotes the matter
action \cite{pois}, and

\begin{equation}\label{kn0}
\kappa_n=\frac{2(n-1)\pi^{n/2} G_{n+1}}{(n-2)(\frac{n}{2}-1)!},
\end{equation}

\noindent is the ($n+1$)-dimensional Einstein coupling constant
\cite{mansori,SMR}. Here,

\begin{equation}\label{G}
G_{n+1}=2 \pi^{1-n/2}\Gamma(\frac{n}{2})\frac{c^3\ell _p^{n-1}
}{\hbar},
\end{equation}

\noindent is the ($n+1$)-dimensional Newtonian gravitational
constant \cite{mann}. In this manner, applying the action
principle to Eq.~(\ref{action}), the Einstein field equations are
achieved as

\begin{equation}\label{kn}
G_{\mu\nu}=\kappa_nT_{\mu\nu}.
\end{equation}

\noindent As a check, for $n=3$, we can easily find

\begin{eqnarray}\label{G3}
&&G_{3+1}=G_4=2 \pi^{-1/2}\Gamma(\frac{3}{2})\frac{c^3\ell
_p^{2}}{\hbar}=\frac{c^3\ell _p^{2}}{\hbar}\equiv G,\nonumber\\
&&\kappa_3=8\pi G\equiv\kappa\\
&&I_G=-\frac{1}{2\kappa}\int R \sqrt{-g}\ d^{4}x,\nonumber
\end{eqnarray}

\noindent which finally leads to $G_{\mu\nu}=\kappa T_{\mu\nu}$.

\subsection*{The Newtonian limit}\label{sub}

For a space with $n$ dimension, the Poisson equation,
corresponding to the Newtonian potential $\phi$ and energy density
$\rho$, is written as \cite{4not,mansori,SMR}

\begin{equation}\label{delphi}
\nabla ^2 \phi=\frac{2G_{n+1}\pi^{n/2} }{(\frac{n}{2}-1)!} \rho.
\end{equation}

\noindent Moreover, the Newtonian limit is evaluated by using the
\cite{4not}

\begin{equation}\label{R00}
R_{00}=\nabla ^2 \phi,
\end{equation}

\noindent relation, for which Eq.~(\ref{kn}) implies \cite{4not}

\begin{equation}\label{Rn}
R_{00}=\left(\frac{n-2}{n-1}\right)\kappa_n \rho,
\end{equation}

\noindent where in accordance with the properties of the Newtonian
limit, the pressure contribution has been ignored \cite{4not}.
Now, combining the above equations with each other, one can easily
reach at Eq.~(\ref{kn0}). Thus, if a $4$-index theory of gravity
is available, then contracting the field equations and by using
the resulting $2$-index field equations, one can find $R_{00}$ and
thus the gravitational coupling constant of the primary $4$-index
theory.

\subsection{$4$-index notation, general remarks and the role of gravitational coupling constant}

Since $R=g^{jl}g^{mn}R_{mjnl}$, by defining the $4$-index metric
$g_{ijkl}$ as $g_{ijkl}=g_{ik}g_{jl}-g_{il}g_{jk}$ which has the
same symmetry as that of the Reimann tensor \cite{4not,4index},
one obtains

\begin{eqnarray}\label{por}
&&R=g^{ik}g^{jl}R_{ijkl}=\frac{1}{n}g^{ik}g^{jl}g_{ijkp}R^p_{\
l}\\&&=\frac{1}{n}g^{ik}g^{jl}g_{ijpl}R^p_{\
k}=\frac{1}{n(n+1)}g^{ik}g^{jl}g_{ijkl}R,\nonumber
\end{eqnarray}

\noindent which helps us in generalizing Eq.~(\ref{lg}) as
\cite{4index}

\begin{equation}\label{lg2}
I_G=-\frac{1}{2\eta_n}\int g^{ik}g^{jl}(A+B+C+D) \sqrt{-g}\
d^{n+1}x,
\end{equation}

\noindent where

\begin{eqnarray}\label{lg20}
&&A=aR_{ijkl}\\&&B=mg_{ijkp}R^p_{\
l}\nonumber\\&&C=mg_{ijpl}R^p_{\
k}\nonumber\\&&D=dg_{ijkl}R\nonumber,
\end{eqnarray}

\noindent and $a$, $m$, and $d$ are unknown Lagrangian
coefficients evaluated later. In Ref~\cite{4index}, for both the
$g_{ijkp}R^p_{\ l}$ and $g_{ijpl}R^p_{\ k}$ terms, author assumed
the same coefficient ($m$). But, since $g_{ijkp}R^p_{\ l}\neq
g_{ijpl}R^p_{\ k}$, their coefficients can be different in
general. We will study some consequences of this case in the next
section.

Now, using Eq.~(\ref{por}), one can see Eq.~(\ref{lg}) is
recovered if we have either

\begin{eqnarray}\label{ein}
&&n[2m+(n+1)d]+a=1\nonumber\\
&&\eta_n=\kappa_n,
\end{eqnarray}

\noindent or

\begin{eqnarray}\label{2}
&&\eta_n\equiv\frac{\kappa_n}{\alpha},
\end{eqnarray}

\noindent where $\alpha\equiv\frac{1}{n[2m+(n+1)d]+a}$. While the
first case claims that the gravitational coupling constant in
$4$-index generalization of Einstein theory ($\eta_n$) is the same
as $\kappa_n$, the second case indicates that $\eta_n$ differs
from $\kappa_n$. Therefore, the definition of gravitational
coupling constant has a key role in getting a $4$-index theory.

Finally, in similarity with the definition of Ricci tensor
($R_{jl}=g^{ik}R_{ijkl}$), and just the same as Ref~\cite{4index},
we assume that there is a $4$-index generalized energy-momentum
tensor $T_{ijkl}$ satisfying the $T_{jl}=g^{ik}T_{ijkl}$
condition, in which $T_{jl}$ is the ordinary $2$-index
energy-momentum tensor representing all sources filling the
background and obtainable by applying the action principle to the
matter Lagrangian, i.e. \cite{4index}

\begin{eqnarray}\label{ma}
&&2\delta I_m=\int T_{jl}\delta g^{jl}\sqrt{-g}\ d^{n+1}x\\&&=\int
T_{ijkl}g^{ik}\delta g^{jl}\sqrt{-g}\ d^{n+1}x\nonumber.
\end{eqnarray}

\subsection*{Action variation and $4$-index theory}

Following the method of Ref~\cite{4index}, the variation of
action~(\ref{lg2}) leads to

\begin{eqnarray}\label{lg3}
&&\delta I_G=-\frac{1}{2\eta_n}\int g^{ik}\big[(m(n-1)+a)\delta
R_{ik}+\delta g^{jl} \big(\nonumber\\&&aR_{ijkl}+m(g_{ijkp}R^p_{\
l}+g_{ijpl}R^p_{\ k})\nonumber\\&& +
(d-\frac{a+2mn+dn(n+1)}{2n})g_{ijkl}R \big)\nonumber\\&&+
(m+nd)\delta(g_{ik}R)\big]\sqrt{-g}\ d^{n+1}x.
\end{eqnarray}

\noindent The integral of $g^{ik}\delta R_{ik}$ will be vanished
\cite{4index}, and the coefficient of the $\delta(g_{ik}R)$ will
be zero whenever $m=-dn$, the primary and simple case studied in
\cite{4index}. In this manner, combining this result with
Eq.~(\ref{ma}), one reaches at

\begin{eqnarray}\label{ne}
&&G_{ijkl}\equiv\big[aR_{ijkl}-nd(g_{ijkp}R^p_{\ l}+g_{ijpl}R^p_{\
k})\nonumber\\&& +
(\frac{dn(1+n)-a}{2n})g_{ijkl}R\big]\nonumber\\&&=\eta_nT_{ijkl}.
\end{eqnarray}

\noindent Now, using Eq.~(\ref{weyl}), we can rewrite this
equation as

\begin{eqnarray}\label{ne2}
&&G_{ijkl}=\big[aC_{ijkl}+F_{ijkl}\big]=\eta_nT_{ijkl},\nonumber\\
&&F_{ijkl}=\frac{a-n(n-1)d}{n-1}(g_{ijkp}R^p_{\ l}+g_{ijpl}R^p_{\
k})\\&&+\frac{n+1}{2n(n-1)}[dn(n-1)-a]g_{ijkl}R.\nonumber
\end{eqnarray}

\subsection*{The Weyl free case}

\noindent The above result indicates that the field equations will
be free of Weyl tensor whenever $a=0$ leading to

\begin{eqnarray}\label{ne3}
&&G_{ijkl}=F_{ijkl}^{(a=0)}=\eta_nT_{ijkl}.
\end{eqnarray}

\noindent In this manner, since we assumed that $T_{jl}$ includes
all sources filling the background, we do not need additional
terms to cancel the Weyl tensor. Additionally, although $F_{ijkl}$
is completely determinable by the Ricci tensor and metric, we
cannot find the $4$-index energy-momentum tensor unless we have
relation between $T_{jl}$ and Ricci tensor meaning that we should
decide about the desired $2$-index theory.

\subsection*{General Relativity}

The contraction of Eq.~(\ref{ne3}) leads to

\begin{eqnarray}\label{ce1}
G_{jl}=\frac{\eta_n}{n(1-n)d}T_{jl},
\end{eqnarray}

\noindent nothing but the Einstein field equations with the
coupling constant $\frac{\eta_n}{n(1-n)d}$. Now, comparing this
equation with~(\ref{kn}), we can easily see that the results
of~(\ref{sub}) is also valid here for
$\frac{\eta_n}{n(1-n)d}=\kappa_n$, a result also compatible with
Eq.~(\ref{2}). Therefore, this analysis cannot give us the values
of $\eta_n$ and $d$ meaning that their values should be evaluated
from other parts. In fact, this analysis shows that the Newtonian
limit and the $d=-\frac{m}{n}$ constraint are enough to recover
the Einstein field equations whenever the $4$-index equations are
Weyl free. As an example, if $\eta_n\equiv\kappa_n$, then we
should have $n(n-1)d=-1$ in full agreement with Eq.~(\ref{ein})
and Ref~\cite{4index}.

In order to find $T_{ijkl}$ and $\eta_n$, we remind that $4$-index
energy-momentum tensor should meet the $T_{jl}=g^{ik}T_{ijkl}$
condition. One can use Eqs.~(\ref{ne3}) and~(\ref{ce1}) in order
to see that only if $n(1-n)d=1$ and

\begin{eqnarray}\label{Tein}
&&T_{ijkl}=\\&&\frac{1}{n-1}(g_{ijkp}T^p_{\ l}+g_{ijpl}T^p_{\
k})-\frac{1}{n(n-1)}g_{ijkl}T,\nonumber
\end{eqnarray}

\noindent then the $T_{jl}=g^{ik}T_{ijkl}$ condition is met in
agreement with Ref~\cite{4index}. In this situation, from the
$\frac{\eta_n}{n(1-n)d}=\kappa_n$ relation, we automatically reach
at $\eta_n=\kappa_n$. Moreover, inserting the above results into
Eq.~(\ref{ne3}), we easily get \cite{4index}

\begin{eqnarray}\label{rev0}
&&B_{ijkl}\equiv
F_{ijkl}^{(a=0,n(1-n)d=1)}=-\frac{n+1}{2n(n-1)}g_{ijkl}R\nonumber\\&&+
\frac{1}{n-1}(g_{ijkp}R^p_{\ l}+g_{ijpl}R^p_{\ k})=\kappa_n
T_{ijkl}.
\end{eqnarray}

\noindent which meets \cite{4index}

\begin{eqnarray}\label{rev1}
G_{jl}=g^{ik}B_{ijkl}.
\end{eqnarray}

\noindent We see that the $T_{jl}=g^{ik}T_{ijkl}$ condition
together with the Newtonian limit automatically give us the value
of $d$ leading to $\eta_n=\kappa_n$. Now, since the
energy-momentum conservation law is met by $T_{jl}$ in Einstein
theory, one can obtain

\begin{eqnarray}\label{div1}
&&\nabla_iT^i_{\
jkl}=\\&&\frac{1}{n-1}\big[(T_{jl;k}-T_{jk;l})-\frac{1}{n}(g_{jl}T_{,k}-g_{jk}T_{,l})\big].\nonumber
\end{eqnarray}

\noindent Bearing the Einstein field equations and Eq.~(\ref{ne3})
in mind and using the above result, one finds

\begin{eqnarray}\label{div2}
&&\nabla_iT^i_{\ jkl}=\frac{1}{\kappa_n}\nabla_iG^i_{\
jkl}=\\&&\frac{1}{\kappa_n(n-1)}\big[(R_{jl;k}-R_{jk;l})+\frac{1}{2n}(g_{jk}R_{,l}-g_{jl}R_{,k})\big],\nonumber
\end{eqnarray}

\noindent which is not always zero \cite{4index}. Therefore, this
study shows that while there is no non-minimal mutual interaction
between the geometry and matter fields in $2$-index general
relativity (or equally $T^i_{\ j;i}=0$), its $4$-index
generalization admits a non-minimal coupling between them meaning
that the divergence of the $4$-index energy-momentum tensor is not
always zero.
\subsection{Some notes on the $m\neq-nd$ case}\label{som}

Here, we want to address the results of the $m\neq-nd$ case. Let
us focus on the last term of Eq.~(\ref{lg3}). In fact, whenever
$m\neq-nd$, then since $g^{ik}g_{ik}=n+1$, we have

\begin{eqnarray}\label{nlg31}
&&g^{ik}\delta(g_{ik}R)=\delta(g^{ik}g_{ik}R)-(\delta
g^{ik})Rg_{ik}\nonumber\\&&=(n+1)\delta R-Rg_{jl}\delta g^{jl},
\end{eqnarray}

\noindent Now, because $\delta R\rightarrow G_{jl}\delta
g^{jl}=g^{ik}B_{ijkl}\delta g^{jl}$, and
$g_{jl}=\frac{1}{n}g_{ijkl}g^{ik}$, the last term of
Eq.~(\ref{nlg31}) leads to

\begin{eqnarray}\label{nterm}
g^{ik}\delta(g_{ik}R)=g^{ik}[(n+1)B_{ijkl}-R\frac{1}{n}g_{ijkl}]\delta
g^{jl},
\end{eqnarray}

\noindent apart of the usual surface term which is the result of
the $\delta R$ term and will be zero at infinity
\cite{4index,4not,pois}. In this manner, bearing Eq.~(\ref{weyl})
in mind, one finally reaches at

\begin{eqnarray}\label{ne00}
&&G_{ijkl}\equiv\big[aC_{ijkl}+\mathbb{F}_{ijkl}\big]+(m+nd)(n+1)B_{ijkl}\nonumber\\&&=\eta_nT_{ijkl},\\
&&\mathbb{F}_{ijkl}=(m+\frac{a}{n-1})(g_{ijkp}R^p_{\
l}+g_{ijpl}R^p_{\
k})\nonumber\\&&-(\gamma+\frac{a}{n(n-1)})g_{ijkl}R.\nonumber\\
&&\gamma\equiv\frac{(dn+2m)(1+n)+a}{2n}.\nonumber
\end{eqnarray}

\noindent Now, contracting this equation
($g^{ik}G_{ijkl}=\eta_ng^{ik}T_{ijkl}=\eta_nT_{jl}$), we find out

\begin{eqnarray}\label{ne200}
&&G_{jl}+\Xi_n\lambda Rg_{jl}=\Xi_nT_{jl},\nonumber\\
&&\lambda=\frac{m(n+1)-2\gamma n+a}{2\eta_n},\\
&&\Xi_n=\frac{\eta_n}{a+n(m+nd(n+1))},\nonumber
\end{eqnarray}

\noindent leading to $R(4\Xi_n\lambda-1)=\Xi_n T$, the trace of
field equations. The above field equations are indeed the Rastall
field equations in which $\lambda$ and $\Xi_n$ denote the Rastall
constant and Rastall gravitational coupling constant, respectively
\cite{rastall}. It is also worthwhile mentioning that the $T^i_{\
j;i}=0$ condition is not met by the Rastall theory claiming that
the geometry and matter fields are coupled with each other in a
non-minimal way \cite{rastall,cmc}. It should be noted that since
we face with Rastall field equations in which $\Xi_n$ is the
gravitational coupling constant, replacing $\kappa_n$ with $\Xi_n$
in Eq.~(\ref{2}), one can reach the above results. $\lambda$ and
$\Xi_n$ are also connected to each other by considering the
Newtonian limit of the Rastall theory \cite{rastall}as
$\frac{\Xi_n}{4\Xi_n\lambda-1}(3\Xi_n\lambda-\frac{1}{2})=\frac{\kappa_n}{2}$.
This result can also be obtained by following the recipe
introduced in~(\ref{sub}). It finally leads to
$\eta_n=\frac{4m-n(8\gamma+d(n+1))+2a}{10m-n(12\gamma+d(n+1))+5a}\kappa_n$
which clearly indicates that we do not have always
$\eta_n=\kappa_n$ \cite{rastall}. The Newtonian limit indeed helps
us in finding relation between $\Xi_n$ with $G$.

It is also easy to check that for $m=-nd$, we get $\lambda=0$ and
thus the Einstein field equations are recovered (for which
$\Xi_n=\kappa_n$), as a desired result in full agreement with
previous achievements. Therefore, in this manner, the contraction
of the $4$-index field equations automatically presents a mutual
interaction between geometry and matter field, i.e. $T_{\mu\nu}^{\
;\nu}=\lambda R_{,\mu}$.

Field equations~(\ref{ne00}) will also become Weyl free whenever
$a=0$. In this manner, bearing the recipe led to Eq.~(\ref{Tein})
in mind, one can use Eq.~(\ref{ne200}) and the trace of field
equations to find $B_{ijkl}$ and $\mathbb{F}_{ijkl}$ which finally
leads to $T_{ijkl}$.
\section{A more general Lagrangian}

Here, after introducing a generalization to Eq.~(\ref{lg20}), we
introduce another $4$-index generalization for Rastall theory
based on the $m=-nd$ case. Some results of the Weyl tensor free
theory are also discussed.

\subsection{Action and its variation}

As we addressed previously, since $g_{ijkp}R^p_{\ l}\neq
g_{ijpl}R^p_{\ k}$, their coefficients can be different in
general. Here, considering different coefficients for these terms,
we are going to study some consequences of the resulting action.
Let us consider a more general form for action~(\ref{lg2}) by
generalizing Eq.~(\ref{lg20}) as

\begin{eqnarray}\label{lg4}
&&A=aR_{ijkl},\\&&B=mg_{ijkp}R^p_{\
l},\nonumber\\&&C=cg_{ijpl}R^p_{\
k},\nonumber\\&&D=dg_{ijkl}R\nonumber,
\end{eqnarray}

\noindent where $c$ (the same as $a$, $m$, and $d$) is an unknown
Lagrangian coefficient. In this situation, Eqs.~(\ref{ein})
and~(\ref{2}) are modified as

\begin{eqnarray}\label{nein}
&&n[m+c+(n+1)d]+a=1\nonumber\\
&&\eta_n=\kappa_n,
\end{eqnarray}

\noindent and

\begin{eqnarray}\label{n2}
&&\eta_n\equiv\frac{\kappa_n}{\beta},
\end{eqnarray}

\noindent in which $\beta\equiv\frac{1}{n[m+c+(n+1)d]+a}$,
respectively. The action variation also leads to

\begin{eqnarray}\label{nlg3}
&&\delta I_G=-\frac{1}{2\eta_n}\int g^{ik}\big[(cn-m+a)\delta
R_{ik}+\delta g^{jl} \big(\nonumber\\&&aR_{ijkl}+mg_{ijkp}R^p_{\
l}+cg_{ijpl}R^p_{\ k}\\&& + (d-\frac{1}{2n\beta})g_{ijkl}R \big)+
(m+nd)\delta(g_{ik}R)\big]\sqrt{-g}\ d^{n+1}x,\nonumber
\end{eqnarray}

\noindent and thus

\begin{eqnarray}\label{nne}
&&G_{ijkl}\equiv\big[aR_{ijkl}-ndg_{ijkp}R^p_{\ l}+cg_{ijpl}R^p_{\
k}\nonumber\\&& +
(\frac{n(d-c)-a}{2n})g_{ijkl}R\big]\nonumber\\&&=\eta_nT_{ijkl},
\end{eqnarray}

\noindent where, as the previous section, we considered the
$m=-nd$ case for which the $\delta(g_{ik}R)$ term is eliminated.
Now, bearing the Weyl tensor~(\ref{weyl}) in mind, we can finally
reach

\begin{eqnarray}\label{nne2}
&&G_{ijkl}=\big[aC_{ijkl}+\mathcal{F}_{ijkl}\big]=\eta_nT_{ijkl},\\
&&\mathcal{F}_{ijkl}=f_1g_{ijkp}R^p_{\ l}+f_2g_{ijpl}R^p_{\
k}+f_3g_{ijkl}R.\nonumber
\end{eqnarray}

\noindent Here,

\begin{eqnarray}\label{co1}
&&f_1\equiv\frac{a-n(n-1)d}{n-1},\
f_2\equiv\frac{a+c(n-1)}{n-1},\nonumber\\
&&f_3\equiv\frac{n(d-c)(n-1)-a(n+1)}{2n(n-1)},
\end{eqnarray}

\noindent and it is easy to see that Eq.~(\ref{ne2}) is recovered
at the appropriate limit of $c=-nd$.

\subsection{Rastall theory}

In the general framework of extended gravity, which has been
discussed in the Introduction of this paper, a renewed interest in
the literature has been recently gained by the theory proposed by
P. Rastall in 1972 \cite{rastall}. In fact, Rastall theory of
gravity presents various good behaviors. It seems consistent with
the Universe age and with the Hubble parameter \cite{Al-Rawaf},
with the helium nucleosynthesis \cite{Al-Rawaf3} and with the
gravitational lensing phenomena \cite{Abdel-Rahman}. It permits an
alternative description for the matter dominated era with respect
to general relativity \cite{Al-Rawaf2}. Such observational
evidences enabled cosmologists to study the various cosmic eras in
the framework of Rastall gravity [34 - 39]. In addition, Rastall
gravity should not present the entropy and age problems of
standard cosmology \cite{Fabris2}. As we previously stressed, the
fundamental issue concerning Rastall gravity is the presence of a
non-divergence-free energy-momentum. For the sake of completeness,
we recall that also the so called curvature-matter non-minimal
theory of gravity shows a similar behavior because also in this
theory the matter and geometry are coupled to each other in such a
way that the ordinary-energy momentum conservation law is not met
\cite{cmc,cmc1,cmc2,Fabris2,Allemandi,Nojiri}.

Now, let us restart our discussion. In general, since
$g^{ik}C_{ijkl}=0$, by contracting Eq.~(\ref{nne2}), we can obtain

\begin{eqnarray}\label{nef}
&&G_{jl}+\Xi_n\lambda Rg_{jl}=\Xi_nT_{jl},
\end{eqnarray}

\noindent as the most general obtainable $2$-index field
equations. As the previous section, $\Xi_n$ and $\lambda$ denote
the Rastall constant and Rastall gravitational coupling constant,
respectively, but here, they are evaluated as

\begin{eqnarray}\label{nef1}
&&\Xi_n\equiv\frac{\eta_n}{a-n^2d-c},\\&&\lambda\equiv\frac{1-n}{2\eta_n}[c+nd].\nonumber
\end{eqnarray}

\noindent Once again, we see that the contraction of $4$-index
theory guides us in general to a non-minimal interaction between
geometry and matter fields in the Rastall way.

We can also easily see that the $c=m=-nd$ case (or equally
$\lambda=0$) leads to the Einstein field equations. Therefore, as
an expected result, the Einstein case can be considered as the
subclass ($c=m=-nd$) of the Lagrangian introduced
here~(\ref{lg4}), and thus this section. The Newtonian limit of
the Rastall theory implies
$\frac{\Xi_n}{4\Xi_n\lambda-1}(3\Xi_n\lambda-\frac{1}{2})=\frac{\kappa_n}{2}$
\cite{rastall}, combined with Eq.~(\ref{nef1}) to obtain
$\eta_n=\frac{(a-n^2d-c)[c(3-2n)+d(2-n)-a]}{c(4-3n)+dn(3-2n)-a}\kappa_n$
indicating that we do not have always $\eta_n=\kappa_n$.
Therefore, even for $m=-nd$, the $4$-index formalism, introduced
in Ref~\cite{4index}, can provide a Lagrangian description for
Rastall theory \cite{rl1,rl2,rl3}.

\subsection*{The Weyl free case ($a=0$)}

In this manner ($a=0$), ane can finally find $T_{ijkl}$ by using
Eq.~(\ref{nef}) and the definition of $\mathcal{F}_{ijkl}$ as

\begin{eqnarray}\label{emc1}
T_{ijkl}=-\frac{[f_1g_{ijkp}T^p_{\ l}+f_2g_{ijpl}T^p_{\
k}-f_4g_{ijkl}T]}{n^2d+c},
\end{eqnarray}

\noindent where

\begin{eqnarray}\label{co2}
f_4\equiv
\frac{1}{4\Xi_n\lambda-1}\big(\frac{2c+n(d-c)}{-2(n^2d+c)}[f_1+f_2]^{a=0}-f_3^{a=0}\big)
\end{eqnarray}

\noindent If we want to obtain the same $T_{ijkl}$ as that of the
Einstein case~(\ref{Tein}), then the Lagrangian coefficients are
constrained by

\begin{eqnarray}\label{co3}
&&f_1^{a=0}=f_2^{a=0}=nf_4=-\frac{n^2d+c}{n-1},
\end{eqnarray}

\noindent leading to $c=-nd$. As we saw in previous section, the
$m=c=-nd$ case can only cover the Einstein field equations.
Therefore, although both the Rastall and Einstein theory has the
same $T_{jl}$~(\ref{ma}), and they can be classified as the
subclasses of one Lagrangian with the same $T_{jl}$ meeting
Eq.~(\ref{ma}), their $4$-index energy-momentum tensor is
different. A difference which is the result of the existence of a
non-minimal mutual interaction between the geometry and matter
fields in Rastall theory.

It is also worthwhile mentioning that the $T_{jl}=g^{ik}T_{ijkl}$
condition is satisfied whenever we have $c=nf_4$ leading to
$c=\frac{dn}{n-2}$. Inserting this result into Eq.~(\ref{nef1}),
one easily finds $\Xi_n=\frac{\eta_n(2-n)}{nd(n(n-2)+1)}$ and
$\lambda=\frac{nd(n-1)^2}{2\eta_n(2-n)}$. This achievement
indicates that the gravitational coupling constant of $4$-index
Rastall theory ($\eta_n$) differs from that of the ordinary
$2$-index Rastall theory ($\Xi_n$), a result in agreement with
subsection~(\ref{som}). Thus, these coupling constants will be the
same only if we have $d=\frac{2-n}{n(n(n-2)+1)}$ leading to
$c=\frac{1}{n(2-n)-1}$.

Just the same as the Einstein case, Eq.~(\ref{emc1}) indicates
that we have not always $\nabla_iT^i_{\ jkl}=0$, meaning that
there is also a non-minimal coupling between the geometry and
matter fields in $4$-index generalization of Rastall theory, a
result in accordance with the Rastall hypothesis \cite{rastall}.
In fact, since $T^i_{\ j;i}=\lambda
R_{,j}=\frac{\Xi_n\lambda}{4\Xi_n\lambda-1}T_{,j}$ in Rastall
theory \cite{rastall}, by bearing the definition of
$\mathcal{F}_{ijkl}$ in mind~(\ref{nne2}), and combining the
energy-momentum conservation law of Rastall theory with
Eqs.~(\ref{nne2}) and~(\ref{emc1}), one can easily check that

\begin{eqnarray}\label{emc2}
&&\nabla_iG^i_{\ jkl}=\eta_n\nabla_iT^i_{\ jkl}.
\end{eqnarray}

\noindent Indeed, since $a=0$ leading to
$\mathcal{F}_{ijkl}=\eta_nT_{ijkl}$, the validity of this result
has been guaranteed. Therefore, just the same as the original
Rastall theory \cite{rastall}, the energy-momentum conservation
law is not always satisfied in the $4$-index generalization of
this theory.

\section{Conclusion}

In summary, our survey shows that although the $d=-\frac{m}{n}$
constraint is sufficient to recover the field equations of general
relativity (Eqs.~(\ref{rev1}) and~(\ref{ce1})), the fulfilment of
the $T_{jl}=g^{ik}T_{ijkl}$ requirement necessitates
$\eta_n=\kappa_n$. It has been obtained that $\nabla_iT^i_{\
jkl}\neq0$ in general, meaning that a non-minimal coupling between
geometry and matter fields is automatically allowed in $4$-index
approach, a result independent of the divergence amount of
energy-momentum tensor ($T_{jl}^{\ \ ;l}$). In addition, we also
found out that the contraction of the $4$-index field equations
may in general bring us to a non-minimum interaction between
geometry and matter field. It has also been obtained that, unlike
the Einstein case, the gravitational coupling constant of
$4$-index Rastall theory ($\eta_n$) generally differs from that of
the ordinary $2$-index Rastall theory ($\Xi_n$).

Besides, we addressed some general expressions for $4$-index
energy-momentum tensor, depending on the values of the Lagrangian
coefficients, which can reduce to those of the Einstein and
Rastall cases by choosing their related coefficients. Therefore,
it should be noted that the $4$-index energy-momentum tensor of
Rastall differs from that of Einstein, while $T_{jl}$ is evaluated
from a unique action principle in both theory (see Eq.~(\ref{ma})
and Ref~\cite{rl5}). This difference is in line with the
difference between the $2$-index Rastall and Einstein theories.

We also found out that Rastall field equations can be obtained
from Lagrangian~(\ref{lg2}) if $m\neq-nd$, a result claiming that
both the Rastall and Einstein theory are subclasses of one general
Lagrangian. On the other hand, in the third section, generalizing
the Lagrangian~(\ref{lg2}), we could again get a $4$-index
generalization for the Rastall theory and thus a Lagrangian
description of this theory. Moreover, although in our calculations
$m=-nd$, we saw that the results of section ($\textmd{II}$) can be
obtained as the special case ($c=-nd$) of this section. This
result also indicates that both the Einstein and Rastall theory
can be considered as the subclasses of one $4$-index theory.

It is worthwhile to remind that $i$) the Riemann tensor has a
crucial role in the Riemannian geometry, and $ii$) the key point
in writing the Lagrangian~(\ref{lg2}) is Eq.~(\ref{por}) giving us
the some possibilities of writing $R$ using the $g^{ik}$ and
$g_{ijkp}$ metrics together with the Riemann tensor and its
contracted form (the Ricci tensor). In the Riemannian geometry,
there are also another curvature invariants built by the Reimann
tensor such as the Kretschmann scalar \cite{4not} and the
Carminati-McLenaghan invariants \cite{CMI}. Hence, writing $R$ (or
equally, the Lagrangian of the geometrical part of the Einstein
theory) in terms of other invariants, and by using the $g_{ijkp}$
notion, one may get another Lagrangians instead of~(\ref{lg2}). In
this manner, one may find another $4$-index field equations which
should cover the Einstein field equations after contraction, and
also the Newtonian gravity after taking the weak field limit. It
was not our aim to study such possibilities, and can be considered
as an interesting subject for the future works.

\section*{Acknowledgments}
We are grateful to the anonymous reviewer for worthy hints and
constructive comments. The work of H. Moradpour has been supported
financially by Research Institute for Astronomy \& Astrophysics of
Maragha (RIAAM).


\end{document}